\documentclass[letterpaper]{article}
\usepackage[draft]{aaai25}
\usepackage{times}
\usepackage{helvet}
\usepackage{courier}
\usepackage[hyphens]{url}
\usepackage{graphicx}
\usepackage{natbib}
\usepackage{caption}
\usepackage{placeins}
\usepackage{multicol}
\usepackage{amsmath,amssymb}
\usepackage{booktabs}
\usepackage{array}
\newcommand{\tablenote}[2][0.95\columnwidth]{%
\vspace{0.45em}
\begin{minipage}{#1}
\scriptsize\emph{Note.}\enspace #2
\end{minipage}}
\title{When Exposure Is Not Attention: Auditing the Preference-Exposure-Consumption Gap in Personalized News Recommenders}
\author{Woojin Park}
\affiliations{Department of Industrial and Management Engineering, Korea University\\
Seoul, Republic of Korea\\
\texttt{woojin\_park@korea.ac.kr}}

\begin{document}
\maketitle

\begin{abstract}
Personalized news platforms are often evaluated as if stated preferences, logged recommendation exposure, and click consumption form a single coherent pipeline. Collapsing these layers can distort audit conclusions: a platform may appear more aligned or diverse than observed click consumption supports, which can misdirect diversity governance or algorithmic intervention. We introduce a reusable Preference-Exposure-Consumption (PEC) audit framework that separates stated preference, observed weighted profile state, logged recommendation exposure, app-surface pathways, and click consumption under explicit observability boundaries. Using six months of logs from a deployed mobile news application, we audit 1,583 user profiles, 95,143 logged recommendation items, and 17,512 click events. Each trace type contributes distinct information; none directly substitutes for another. Preference-consumption alignment exceeds chance-based null baselines but captures only part of users' top-consumed category set. Logged recommendation lists contain clicked articles more often than a date-matched candidate-pool baseline predicts (11.29\% vs 9.13\%; top-5 lift 1.47x), but many clicks arrive through other app surfaces. Raw exposure-consumption diversity gaps shrink under count matching, yet concentration mismatch persists in the audit-eligible cohort (HHI gap 0.082). Together, the results show that audit conclusions change depending on whether platforms measure stated preference, logged exposure, or click consumption.
\end{abstract}

\section{Introduction}

Mobile news platforms have become infrastructures for public attention. They do not merely deliver articles to users; they shape which stories and viewpoints are surfaced within everyday information routines, and these ranking choices embed editorial, commercial, and democratic value judgments (Helberger 2019; DeVito 2017). As a result, personalized news is not only a recommender-system problem. It is also a social computing problem about how digital platforms shape attention, diversity, and information access.

A common technical framing treats personalization as a pipeline. Users first state their interests, the platform uses these interests and subsequent behavior to generate recommendations, and users then consume recommended content. User-centric evaluation distinguishes objective system aspects, perceived system qualities, user experience, and observable interaction (Knijnenburg et al. 2012), while purpose-driven evaluation requires metrics to align with intended recommender purposes and stakeholder goals (Jannach and Adomavicius 2016). This view remains useful, but recent platform audits and feed studies show that ranking systems and network structure can shape bounded exposure patterns without equating exposure with engagement (Bandy and Lazovich 2023; Duskin et al. 2026), while interaction and promotion studies show that platform affordances create additional pathways through which content becomes visible and actionable (An et al. 2024; Schlessinger et al. 2023). In a deployed mobile news application, users may encounter articles through ranked recommendation lists, headline modules, category tabs, newsroom views, home/feed surfaces, search, notifications, and article-detail pathways. A click is therefore not always the endpoint of a single recommendation event. Clicks provide informative but presentation-biased implicit feedback whose interpretation depends on rank and the surrounding choice set (Joachims et al. 2005), and broader digital-trace research cautions that platform-generated behavioral traces should not be treated as complete or transparent measures of human behavior (Lazer et al. 2021). In our logs, a click records an article opening but does not measure dwell time, reading depth, or completion; we therefore interpret it as observed click consumption rather than complete attention.

This distinction matters for both recommender-system evaluation and social computing research. News recommendation should be evaluated beyond relevance and click-based performance because recommender design can implicate diversity, user autonomy, and broader democratic values (Helberger 2019; Vrijenhoek et al. 2021). At the same time, selective-exposure and online-news studies ask whether concentrated information diets arise from algorithmic filtering, homophilous networks, or users' own navigation and attitude-congruent choices (Hart et al. 2009; Flaxman, Goel, and Rao 2016). These perspectives are often studied separately. Recommender-system research may focus on what algorithms expose; selective-exposure research may focus on what users consume. Exposure and consumption are connected but not interchangeable.

We study the gap among three layers of personalized news: what users say they want, what the system logs as recommendation exposure, and what users actually consume. We call this mismatch the Preference-Exposure-Consumption gap, or PEC gap. The PEC gap is not automatically a sign of system failure: its interpretation depends on the platform's intended function, the values at stake, and the form of diversity being evaluated (Helberger 2019). It may reflect user agency, alternative discovery surfaces, interface design, breaking-news salience, habitual browsing, or incomplete exposure logging. Measuring the gap is nevertheless important because it reveals how much each layer of personalization actually explains.

We conduct a platform-internal observational audit of a deployed mobile news application. The dataset covers six months of natural activity and links user profiles, recommendation logs, behavior events, click events, and article metadata. This lets us ask four questions: \textbf{RQ1:} To what extent do stated news preferences align with actual consumption? \textbf{RQ2:} How much observed consumption can be traced to logged recommendation exposure? \textbf{RQ3:} How do exposure-side and consumption-side diversity differ? \textbf{RQ4:} To what extent are preference-consumption gaps structured by observed profile-state features and behavioral usage patterns?

The answers matter because platform evaluation often collapses different traces into a single notion of user preference. A click can be treated as preference, a stated category can be treated as intent, and a logged recommendation record can be treated as exposure. In practice, these are different traces of a broader sociotechnical system. If these layers are collapsed, platforms can overstate preference alignment, overclaim diversity delivery, or choose interventions for the wrong part of the product. The PEC framework gives those traces separate analytical roles and asks where their relationships align, where they weaken, and where they should not be interpreted as direct substitutes for one another.

\begin{figure*}[t]
\centering
\includegraphics[width=0.80\textwidth,trim=10pt 18pt 10pt 2pt,clip]{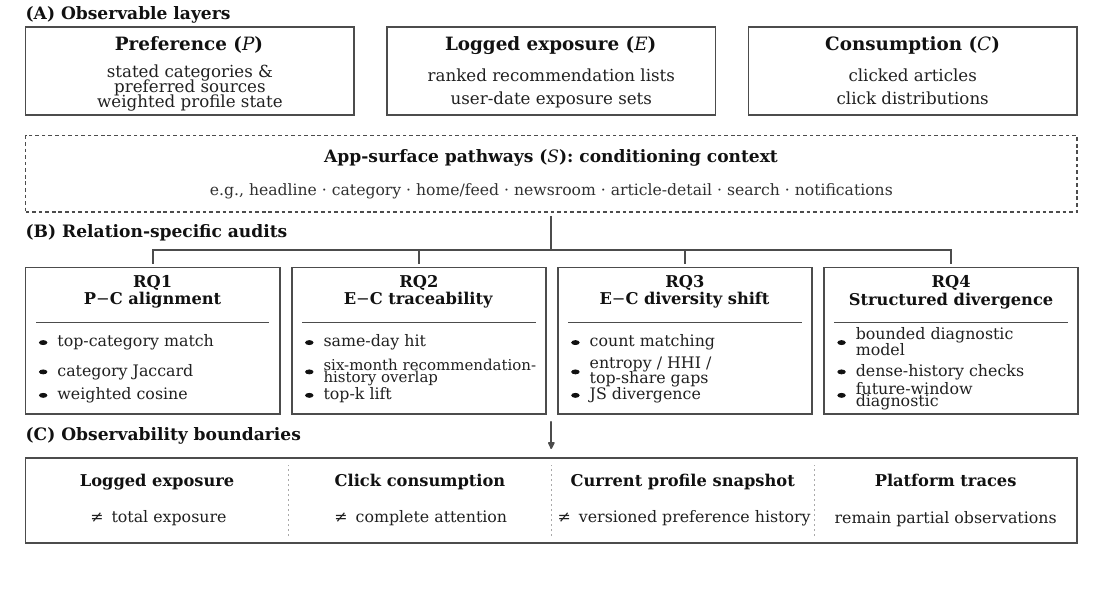}
\caption{PEC audit framework. (A) Preference (P), logged exposure (E), and click consumption (C) are separated as distinct observable layers, with app-surface pathways (S) as conditioning context. (B) Four relation-specific audits examine P--C alignment, E--C traceability, E--C diversity shift, and structured divergence. (C) Audit claims are interpreted under explicit observability boundaries.}
\label{fig:pec-framework}
\end{figure*}

The paper's central claim is that preferences, logged exposure, and clicks are informative but not interchangeable. PEC is therefore not a single composite score; it is a relation-specific audit protocol for comparing stated preference, logged recommendation exposure, and observed consumption. Each comparison is tied to its layer relation, eligible denominator, validity control, and observability boundary. Surface pathways provide interface context rather than a fourth substantive preference, exposure, or consumption signal. Figure~\ref{fig:pec-framework} summarizes the P, E, and C layers, the S conditioning context, and relation-specific audit outputs.

\subsection*{Contributions}

First, we introduce PEC as a reusable relation-specific audit protocol that separates stated preference, observed profile state, logged recommendation exposure, app-surface pathways, and click consumption, with relation-appropriate denominators and observability boundaries. Second, using six months of linked operational logs, we characterize preference-consumption alignment, exposure-consumption traceability, and layer-wise diversity shifts in a deployed multi-surface news application. Third, we show how candidate-pool and popularity-aware baselines, count matching, dense-history cohorts, alternative labels, and future-window validation yield bounded audit claims rather than ranking or causal-effect claims.

\section{Related Work}

News recommendation should be evaluated beyond accuracy or click prediction because news carries civic and informational consequences. Helberger frames news recommenders in relation to democratic roles rather than only individual relevance (Helberger 2019). Jannach and Adomavicius argue that recommender evaluation should align stakeholder purposes, operational tasks, and metrics (Jannach and Adomavicius 2016), and mission-oriented news recommender work treats diversity as a multidimensional, normatively grounded objective (Vrijenhoek et al. 2021). We build on this line by separating observable category- and publisher-level diversity represented in logged recommendation exposure from diversity that users actually consume.

Selective-exposure and online-news studies show that information diets reflect both platform mediation and user choice. Users often prefer congenial information, but network exposure, ranking, search, homophily, and active selection jointly shape what is seen and clicked (Hart et al. 2009; Bakshy, Messing, and Adamic 2015; Flaxman, Goel, and Rao 2016; Robertson et al. 2023). Bakshy, Messing, and Adamic separated potential exposure, ranked News Feed exposure, and click selection; Robertson et al. similarly distinguish search-result exposure from selected engagement. Recommender-diversity simulations and user-centric evaluation further caution that diversity and experience depend on the algorithm, user history, metric, and perceived system qualities (Möller et al. 2018; Knijnenburg et al. 2012). PEC builds on this distinction by incorporating stated preference, profile state, and multi-surface consumption pathways under explicit observability boundaries. We do not make ideological-polarization claims because the present data lack validated ideology labels; instead, we measure category- and publisher-level gaps observable in app metadata.

Algorithm audit and accountability research develops systematic ways to examine opaque platform outputs and the values embedded in prioritization, classification, association, filtering, and feed curation (Sandvig et al. 2014; Diakopoulos 2015; DeVito 2017). PEC adopts this audit orientation, but as a platform-internal observational measurement audit using linked operational traces rather than a black-box experiment that manipulates accounts, recommendations, or ranking outputs.

Recent platform studies motivate separating exposure from consumption. Randomized feed and counterfactual-bot studies show that ranking can redistribute exposure while observed consumption or attitudes need not identify recommender influence by itself (Huszár et al. 2022; Guess et al. 2023; Hosseinmardi et al. 2024). Search and engagement studies likewise show that selected engagement can differ from exposure and from users' item-level stated preferences (Robertson et al. 2023; Milli et al. 2025).

Within ICWSM, prior work has audited news curation, bounded exposure, and interaction pathways in deployed platforms, including Apple News, Twitter timelines, cross-party interaction, follow-network exposure, and trending-topic surfaces (Bandy and Diakopoulos 2020; Bandy and Lazovich 2023; An et al. 2024; Duskin et al. 2026; Schlessinger et al. 2023). These studies show why exposure audits require explicit denominators and why interface or network pathways can shape downstream engagement. PEC complements this line by linking internal preference, recommendation, surface-pathway, and click traces under explicit observability boundaries.

Digital trace research emphasizes that platform logs are constructed, system-dependent, and bounded measures rather than complete records of social activity (Lazer et al. 2021). PEC does not claim that preference, exposure, and consumption have never been distinguished; its contribution is to operationalize these distinctions within a relation-specific audit that assigns each comparison its own denominator, validity control, and observability boundary.

\section{Data and Platform Context}

We analyze six months of logs from a deployed mobile news application, covering January 1 to June 30, 2026. The platform provides personalized recommendations and allows users to consume news through multiple app surfaces. The analysis combines user profile data, recommendation logs, behavior logs, click events, and article metadata. Recommendation records capture ranked lists generated and logged for each user-date; they represent observable recommendation availability rather than verified viewport-level impressions. The recommendation logs represent a logged recommendation layer, not every possible impression across the product. Because recommendation records are available only for observed recommendation user-days, same-day hit is interpreted as traceability within the logged recommendation layer rather than as CTR or total exposure. Because within-day recommendation-click ordering is unavailable, same-day matching indicates temporal co-occurrence in the logged layer, not verified pre-click exposure.

Eligible denominators therefore vary by audit relation: preference-consumption analyses require both profile records and matched clicks, exposure-consumption analyses require comparable click rows with available recommendation lists, and exposure-consumption diversity-shift checks further restrict to users with enough observed clicks and logged exposure to support stable user-level comparisons.

We treat the app as a multi-surface environment. Users may enter articles through recommendation lists, headline-oriented modules, category tabs, newsroom views, home/feed surfaces, search, notifications, article-detail pathways, and other routes. This distinction matters because RQ2 is not asking whether a user could ever have seen an article. It asks whether a click is traceable to the logged recommendation layer available in the data.

Audit context. The dataset consists of operational logs collected during ordinary use of the deployed mobile news application. The analysis was conducted as an internal platform audit using platform-internal identifiers; no direct identifiers are reported. No experimental intervention was performed, and the analysis did not alter users' recommendations, ranking outputs, notifications, or app experience. Access to raw logs was limited to authorized platform personnel for research and auditing purposes.

\begin{table}[!htbp]
\centering
\caption{Dataset components.}
\small
\setlength{\tabcolsep}{4pt}
\renewcommand{\arraystretch}{1.12}
\begin{tabular}{@{}>{\raggedright\arraybackslash}p{0.62\columnwidth}>{\raggedright\arraybackslash}p{0.28\columnwidth}@{}}
\toprule
\textbf{Component} & \textbf{Count / coverage} \\
\midrule
News metadata & 128,385 \\
User profiles & 1,583 \\
Deduplicated behavior events & 471,027 \\
Click events & 17,512 \\
Matched click rows & 17,232 \\
Recommendation rows & 4,613 \\
Logged recommendation items & 95,143 \\
Matched recommendation items & 90,076 \\
\bottomrule
\end{tabular}
\end{table}

Behavior events were deduplicated using timestamp, user identifier, session identifier, event name, news identifier, and date keys. Recommendation lists were exploded into item-level records containing user, date, rank, and news identifier. Click events were joined with article metadata to obtain category, publisher, headline, tag, summary, and breaking-news fields used to characterize clicked content.

User profiles contain two preference representations used for distinct alignment checks. The first is an explicit user-selection layer for categories and preferred sources, which we use for set-based top-category and Jaccard alignment. The second is a weighted profile-state distribution over categories and sources. These weighted fields are initialized during mandatory onboarding and can later be manually revised by users through the preference panel. The available user table stores one current profile snapshot per user and does not preserve a complete edit history of preference revisions. We therefore treat weighted preferences as observed user-maintained profile-state distributions rather than as time-stamped onboarding-only preferences.

\begin{figure}[t]
\centering
\includegraphics[width=\columnwidth,trim=7pt 7pt 7pt 6pt,clip]{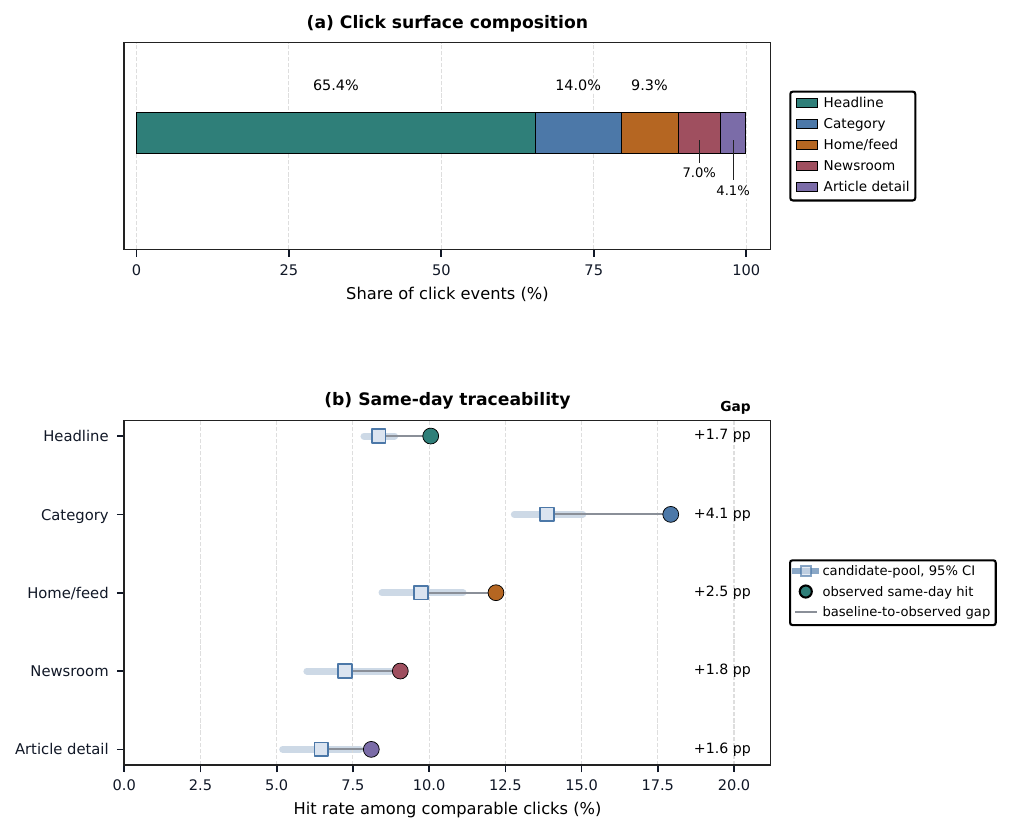}
\caption{Multi-surface click traceability: click-surface composition and observed same-day hit rates against date-matched logged-list candidate-pool baselines.}
\label{fig:surface-traceability}
\end{figure}

For surface analysis, each matched click is assigned exactly one entry pathway using the final recorded click-time metadata. The priority rule uses notification type, source-path metadata, section labels, and headline flags to distinguish notification, headline, category, home/feed, article-detail, newsroom, search, podcast, insights, and other/unknown pathways. Missing or invalid surface metadata are retained as other/unknown in aggregate counts; surface-level hit rates use clicks for which a same-day recommendation list is available as the denominator. Fifteen other/unknown matched clicks are retained in aggregate totals but omitted from the five major-surface display in Figure~\ref{fig:surface-traceability} and Table 8.

\section{Measurement Framework}

PEC audit protocol. A PEC audit separates platform traces into three substantive observable layers: preference (P), logged exposure (E), and consumption (C), with surface pathways (S) represented as conditioning context. Surface pathways describe interface entry routes for clicked articles; the analysis compares relations across pathways without estimating causal mediation. The layers and surface context link through user, item, time, and surface keys, without substituting for one another. The audit computes four families of quantities: preference-consumption alignment, exposure-consumption traceability, exposure-consumption diversity shift, and structured divergence. Beyond observing that layers differ, a PEC audit specifies the audited relation, eligible denominator, validity control, and interpretation permitted by available traces. Each claim is paired with explicit observability limits. PEC is an audit protocol, not a ranking objective: it defines which traces can be compared and which claims the available logs cannot support.

\begin{table*}[!t]
\centering
\caption*{\textbf{Protocol 1: PEC audit protocol}}
\small
\setlength{\tabcolsep}{4pt}
\renewcommand{\arraystretch}{1.12}
\begin{tabular}{@{}p{0.08\textwidth}p{0.18\textwidth}p{0.49\textwidth}p{0.18\textwidth}@{}}
\toprule
\textbf{Phase} & \textbf{Layer / relation} & \textbf{Operation} & \textbf{Audit role} \\
\midrule
1 & P: Preference & Build stated-preference sets and weighted profile-state distributions. & Preference layer \\
1 & E: Logged exposure & Build user-date recommendation sets, ranks, and exposure unions. & Exposure layer \\
1 & S: Surfaces & Assign click-time entry pathways from interface metadata. & Surface-context layer \\
1 & C: Consumption & Build clicked-item sets and click distributions linked to article metadata. & Consumption layer \\
2 & P $\leftrightarrow$ C & Compute top-category match, Jaccard overlap, cosine similarity, and JS divergence. & Preference-consumption alignment \\
2 & E $\leftrightarrow$ C & Trace clicks to same-day lists, exposure history, candidate-pool baselines, and popularity-aware baselines. & Exposure-consumption traceability \\
2 & E $\leftrightarrow$ C & Count-match exposure to clicks; compute entropy, HHI, top-share, effective categories, and JS divergence. & Diversity shift \\
2 & P, E, C & Fit bounded diagnostic models excluding direct alignment quantities. & Structured divergence \\
3 & Output & Pair each empirical claim with its observability boundary. & Audit report \\
\bottomrule
\end{tabular}
\tablenote[0.93\textwidth]{Protocol 1 is a measurement protocol, not a ranking objective or production targeting procedure. It separates what is observed, what is compared, and what remains outside the observable scope of the logs.}
\end{table*}

\subsection{The PEC Gap}

For each user, we define four audit objects in ordinary log terms: stated category and source preferences, user-date logged recommendation exposure, clicked articles linked to article metadata, and user-level profile, exposure, and click distributions. The user's top-clicked-category set contains the top three clicked categories, or fewer when fewer clicked categories are observed. Each article has a category, publisher, and date. We define the PEC gap as a mismatch among these preference, exposure, and consumption layers. Table 2 renders the corresponding metric equations.

Preference-consumption alignment is measured in two ways. For explicit selections, we use top-category match and category-level Jaccard overlap. For observed weighted profile-state distributions, we compute the profile weight assigned to a user's top clicked category, the total preference mass assigned to the top clicked categories, cosine similarity between the preference and click distributions, and Jensen-Shannon divergence between the two distributions. Exposure-consumption traceability is measured using same-day hit, six-month recommendation-history overlap, and top-k lift over a date-matched logged-list candidate-pool baseline. Diversity is measured using normalized Shannon entropy, top-category share, HHI concentration, effective number of categories, and user-level Jensen-Shannon divergence.

These metrics intentionally operate at category and publisher levels rather than ideological or stance levels. This keeps the claims aligned with available metadata. Category-level diversity is not equivalent to viewpoint diversity, but it is still relevant for a mobile news product because categories structure the interface, stated category preferences, recommendation outputs, and many user navigation paths.

For count-matched diversity checks, we repeatedly downsample each user's logged recommendation exposure to the user's number of observed clicks and recompute the metric. We orient entropy, HHI, and top-share gaps so that positive values indicate narrower observed consumption relative to matched exposure: exposure minus click for entropy, but click minus exposure for HHI and top-share. HHI concentration is the sum of squared category shares; effective categories is the inverse HHI; top-share is the largest category share. Equal-weight, base-2 Jensen-Shannon divergence follows Lin's entropy-based formulation and is computed between each user's logged-exposure and click-consumption category distributions over the union of observed categories (Lin 1991). Because the mixture distribution assigns positive mass wherever either input distribution has positive mass, JS divergence is finite without additive smoothing.

We report two comparable-observation cohorts to avoid treating unstable traces as equally reliable. The click $\geq$5 diversity cohort contains users with at least five matched clicks. The audit-eligible core cohort additionally requires at least 20 matched recommendation items, at least two active click days, and at least two stated preference categories. These criteria are fixed before examining metric values and are used to test whether PEC signals persist among users for whom preference, exposure, and consumption are all sufficiently observed. The main analysis therefore reports relation-specific denominators: 1,583 observed profiles, 698 users eligible for P-C alignment, 553 users with matched exposure-click diversity records, 284 users eligible for count-matched diversity analysis with at least five matched clicks, and 251 users in the audit-eligible diversity core. Modeling cohorts are reported separately in Tables 9--10.

For the top-k logged-list candidate-pool baseline, we preserve each user-date's top-k recommendation-list size. The expected hit rate is computed against the same-date top-k recommendation candidate pool: if a clicked article appears in that date's candidate pool, its baseline probability is the user-date list size divided by the candidate-pool size; otherwise it is zero. Operationally, the candidate pool is the union of articles appearing in logged recommendation lists on the same date after the same metadata-availability filters used for comparable clicks; it is not the full news corpus. Confidence intervals for user-level diversity use user-level bootstraps; row-level traceability intervals are descriptive checks over comparable click rows.

\begin{table*}[t]
\centering
\caption{Metrics used to operationalize the PEC audit.}
\small
\setlength{\tabcolsep}{5pt}
\renewcommand{\arraystretch}{1.22}
\begin{tabular}{@{}p{0.24\textwidth}p{0.68\textwidth}@{}}
\toprule
\textbf{Audit quantity} & \textbf{Operational equation} \\
\midrule
Category Jaccard & $J_u=\dfrac{|P^{cat}_u \cap C^{top}_{cat}(u)|}{|P^{cat}_u \cup C^{top}_{cat}(u)|}$ \\
Divergence & $D_u=1-J_u$ \\
Preference-click cosine & $A_u=\cos\!\left(p^{cat}_{u,P},\,p^{cat}_{u,C}\right)$ \\
Same-day hit & $H_{ui}=\mathbf{1}\{i\in E(u,\operatorname{date}(i))\}$ \\
Top-$k$ lift & $\mathrm{Lift}_k=\dfrac{\mathbb{E}[H^{k}_{obs}]}{\mathbb{E}[H^{k}_{base}]}$ \\
Normalized entropy & $H_{\mathrm{norm}}(X)=-\sum_x p_x\log(p_x)\,/\,\log|X^+|,\quad X^+=\{x:p_x>0\}$ \\
HHI / top-share & $\mathrm{HHI}(X)=\sum_x p_x^2,\qquad \mathrm{TopShare}(X)=\max_x p_x$ \\
JS divergence & $\mathrm{JS}_2(p,q)=\tfrac{1}{2}\mathrm{KL}_2(p\parallel m)+\tfrac{1}{2}\mathrm{KL}_2(q\parallel m),\quad m=\tfrac{p+q}{2}$ \\
Count-matched narrowing & $\Delta H_{\mathrm{norm}}=H_{\mathrm{norm}}(E'_u)-H_{\mathrm{norm}}(C_u)$; $\Delta\mathrm{HHI}=\mathrm{HHI}(C_u)-\mathrm{HHI}(E'_u)$; $\Delta\mathrm{TopShare}=\mathrm{TopShare}(C_u)-\mathrm{TopShare}(E'_u)$, with $|E'_u|=|C_u|$ \\
\bottomrule
\end{tabular}
\tablenote[0.93\textwidth]{$C^{top}_{cat}(u)$ denotes each user's top three clicked categories, or fewer when fewer are observed. $E'_u$ is sampled from the user's logged-exposure union. Entropy uses the nonzero support within each sampled distribution, with one-label distributions assigned entropy 0. JS divergence uses equal weights and base-2 logs over union support; zero-mass terms contribute zero in the KL components against the mixture distribution. We report divergence, not its square-root distance.}
\end{table*}

These definitions are not intended to form a new ranking objective. They operationalize the PEC audit by making explicit which layer is being compared, what can be traced in the logs, and where interpretation is limited by observability.

\subsection{Modeling High-Divergence Users}

We use classification diagnostically, not as a recommendation or targeting model, to test whether preference-consumption divergence is associated with observable profile-state and behavioral features rather than random noise or sparse histories.

Using the divergence score $D_u$ defined in Table 2, users above the sample median are labeled high-divergence. Random Forest and Extra Trees classifiers capture nonlinearities and higher-order interactions in tabular predictors without prespecified functional forms (Breiman 2001; Geurts, Ernst, and Wehenkel 2006).

The main same-window model classifies high-divergence users as an audit exercise rather than forecasting future behavior. Weighted-profile predictors include only distributional properties of the observed profile snapshot, such as preference entropy, maximum preference weight, and salient preference count. Full audit feature sets retain aggregate behavioral summaries such as click count, active days, timing, and surface use, but exclude clicked-item and clicked-category identities, clicked-category weights, preference-click cosine, and preference-click divergence.

Because complete preference-edit histories are unavailable, same-window models using weighted profiles are interpreted as audit classifications rather than prospective forecasts. As a future-window diagnostic, we predict April-June high-divergence labels from January-March behavior and exposure features for users active in both windows; this check excludes current weighted-profile snapshot features because historical profile snapshots are not timestamped. The preference component of the April-June label is therefore derived from the available non-versioned profile snapshot, so this diagnostic separates earlier behavior and exposure features from later click outcomes but is not a fully prospective prediction from a pre-period preference state. Performance is estimated using user-level five-fold cross-validation, ensuring that each user appears in only one fold. Across these settings, the model is diagnostic rather than operational: it tests whether the PEC gap is structured by measurable platform behavior.

\section{Results}

We organize the results around four claims: preference representation matters; exposure traceability is partial and surface-dependent; diversity gaps are measurement-layer dependent; and divergence is associated with observable user features. The first two ask how preference and exposure traces align with clicks; the latter two test whether diversity and divergence patterns survive denominator, count, cohort, and label checks.

\subsection{RQ1: Stated Preferences Are Informative but Incomplete}

Among 698 users with profile data and matched click events, stated preferences identify dominant categories but not the full top-consumed category set: the top consumed category appears in stated interests for 77.22\% of users. Treating 19 observed empty stated-category sets as $J_u=0$, primary mean Jaccard is 0.2283 (median 0.2000); restricting to 679 nonempty sets yields a similar mean of 0.2347. A chance-based null check supports this interpretation: shuffled-set and size-preserving category-frequency nulls yield 61.54\% and 54.81\% top-category match and mean Jaccard 0.161 and 0.137, below the observed values. We do not interpret this as profile failure. Instead, stated preference captures dominant orientation better than the breadth, context, and situational variation of actual news consumption.

The two alignment metrics support different interpretations. If the analysis only used top-category match, it would suggest that profiles are broadly aligned. If it only used Jaccard overlap, it would suggest that profiles explain little of the consumption distribution. Together, they show that stated preferences identify dominant categories but do not capture the broader set and situational variation of click behavior. This is the kind of layer divergence the PEC framework is designed to expose.

The observed weighted profile-state distributions sharpen this interpretation. Across 698 users, the mean profile weight assigned to the user's top clicked category is 0.143, and the mean preference mass assigned to the user's top three clicked categories is 0.264. The mean category preference-click cosine is 0.452, indicating moderate directional alignment, while the mean base-2 JS divergence is 0.579, indicating substantial distributional mismatch. These values show that user-maintained weighted profiles carry real signal while leaving substantial unexplained consumption breadth.

Source-level alignment is weaker, though this comparison must be read with support size in mind. Publisher/source space is much larger and sparser than the category taxonomy, so the absolute magnitudes are not directly comparable to category-level weights. In the available source-profile representation, clicked publishers receive little profile mass on average: the mean profile weight assigned to the top clicked publisher is 0.014, and the mean publisher preference-click cosine is 0.080.

\begin{figure}[t]
\centering
\includegraphics[width=\columnwidth,trim=7pt 7pt 7pt 6pt,clip]{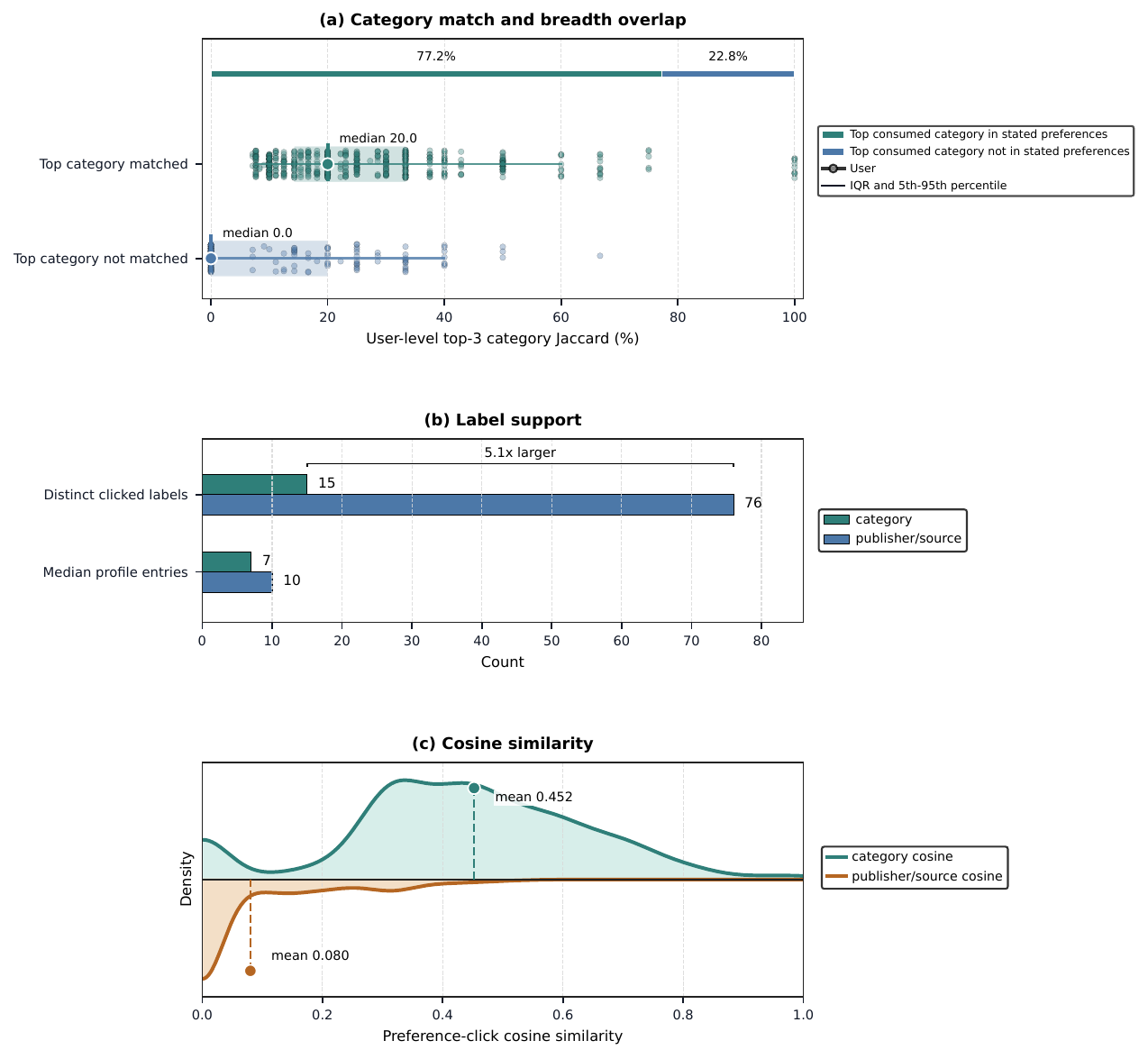}
\caption{Preference-consumption diagnostics: category recovery, Jaccard overlap (empty stated sets contribute $J_u=0$), label support, and category versus source cosine similarity.}
\label{fig:preference-alignment}
\end{figure}

The appendix reports the full tabular summary for this alignment check.

\subsection{RQ2: Only a Small Share of Clicks Is Traceable to Logged Recommendation Exposure}

Among 17,232 matched clicks, 11,440 have comparable same-day recommendation lists. Of these, 11.29\% appear in the same-day list, so logged recommendation records trace only a minority of observed click consumption. The paired observed-minus-baseline difference is 2.16 percentage points under user-clustered bootstrap (95\% CI [0.64, 4.37]). Six-month recommendation-history overlap is 14.37\% of all matched clicks; this historical-union diagnostic relaxes click-date ordering. Neither value is a conventional CTR or direct quality metric; both are traceability statistics for the logged recommendation layer. The same-day rate is timing bounded: it exceeds a date-matched logged-list candidate-pool baseline, while most clicks occur through headline, category, home/feed, newsroom, and article-detail pathways, with only a small remainder on other observed surfaces.

The same-day hit rate is not an evaluation of recommender quality or item attractiveness. Logged recommendation exposure is one traceable pathway among several, not a complete representation of the user's information diet. Evaluating personalization through a single logged list can therefore miss much of the user's path to news.

The same-day hit rate also exceeds simple popularity-aware baselines reported in Appendix D, including a leave-user-out same-day recommendation-popularity baseline (9.52\%) and prior-day click popularity (8.85\%). This comparison is motivated by recommender-systems work showing that recommendation algorithms can over-represent popular items and concentrate exposure in the short head of the catalog (Abdollahpouri, Burke, and Mobasher 2019). We therefore treat popularity as a descriptive non-personalized baseline rather than as a full causal decomposition. The margins are modest and do not isolate editorial salience, so we interpret this as additional traceability beyond the candidate-pool and generic popularity baselines rather than as a causal personalization effect.

\subsection{Surface Pathways}

To characterize the interface pathways through which matched click consumption occurs, we classify matched clicks by available surface metadata. Headline-oriented surfaces account for 65.4\% of matched clicks, category surfaces for 14.0\%, home/feed surfaces for 9.3\%, newsroom surfaces for 7.0\%, and article-detail pathways for 4.1\%. This supports the interpretation that consumption is distributed across multiple surfaces rather than being fully explained by a single recommendation list.

Surface-level traceability also varies. Category clicks have a same-day recommendation hit rate of 17.93\%, while headline clicks have a hit rate of 10.05\%. Different interface pathways therefore connect to the recommendation layer in different ways.

Surface pathways also differ in preference alignment. Among user-surface groups with at least two clicks, category-surface clicks have a higher mean weight on the top clicked category (0.195) than headline clicks (0.122), while home/feed clicks show the highest mean preference-click cosine among major surfaces (0.484). Surface pathways are therefore not merely descriptive labels in this audit; they are associated with differences in how profile state, logged exposure, and click consumption align.

Rank-aware traceability provides a second view of the recommendation signal. The top-5 logged recommendation list contains the clicked article for 4.99\% of comparable clicks, compared with an exact date-matched candidate-pool expectation of 3.41\% (1.47x lift). At top-20, the hit rate rises to 11.05\% with 1.28x lift. Thus, the logged recommendation layer is not irrelevant; it carries measurable signal even though it does not exhaust observed click consumption.

The signal is also surface-dependent. Category-surface clicks show the strongest lift among major pathways: actual same-day hit 17.93\% versus candidate-pool expectation 13.86\% (1.29x). This supports the multi-surface interpretation: different pathways have different relationships to logged recommendation exposure.

\subsection{RQ3: Diversity Gaps Depend on the Measurement Layer}

In the raw logs, exposure diversity and click-consumption diversity are layer-specific quantities: pooled recommendation-category entropy is 0.8352, while pooled click-category entropy is 0.5401. Recommendation top-category share is 0.3704, while click top-category share is 0.6928. Exposure diversity should therefore not be confused with observed consumption diversity.

Viewed alone, recommendation logs make the platform appear relatively diverse, while click logs make consumption appear narrower. Viewed together, the layers show that diversity can narrow between logged exposure and observed clicks without assigning responsibility solely to the algorithm or solely to the user. Algorithmic curation, interface pathways, and selective consumption operate at different layers of the same process.

Exposure-side diversity is diversity represented in the logged recommendation layer; click-side diversity is observed click-consumption diversity, not complete attention. A platform may show the first without observing the second; consistent with mission-oriented diversity work, diversity claims should specify the layer at which diversity is measured (Vrijenhoek et al. 2021).

The same narrowing appears at the publisher level: pooled recommendation-publisher entropy is 0.9129, while pooled click-publisher entropy is 0.6256. Separately, a user-level bootstrap confirms a mean user-level publisher entropy gap of 0.274, with 95\% CI [0.238, 0.309].

We next test whether the entropy gap is mechanically produced by users having many more recommendation exposures than clicks. We repeatedly downsample each user's logged exposure to the user's click count and recompute entropy. Among users with at least five clicks, the count-matched category gap remains positive but much smaller: 0.032, 95\% CI [0.013, 0.054]. Among users with at least ten clicks, the category gap is near zero (0.010), and publisher narrowing is not robust after count adjustment. The raw diversity gap is therefore partly a sparse-observation effect, not pure behavioral narrowing.

Because entropy is sensitive to sparse samples, we also use count-matched concentration metrics. In the audit-eligible core cohort, click consumption remains more concentrated than matched exposure (HHI gap 0.082, 95\% CI [0.064, 0.101]; top-share gap 0.069, 95\% CI [0.049, 0.089]), while base-2 category JS divergence remains high (0.471, 95\% CI [0.449, 0.494]). RQ3's robust signal is count-adjusted concentration and distributional mismatch, not pure entropy narrowing.

\begin{figure}[t]
\centering
\includegraphics[width=\columnwidth,trim=7pt 8pt 7pt 6pt,clip]{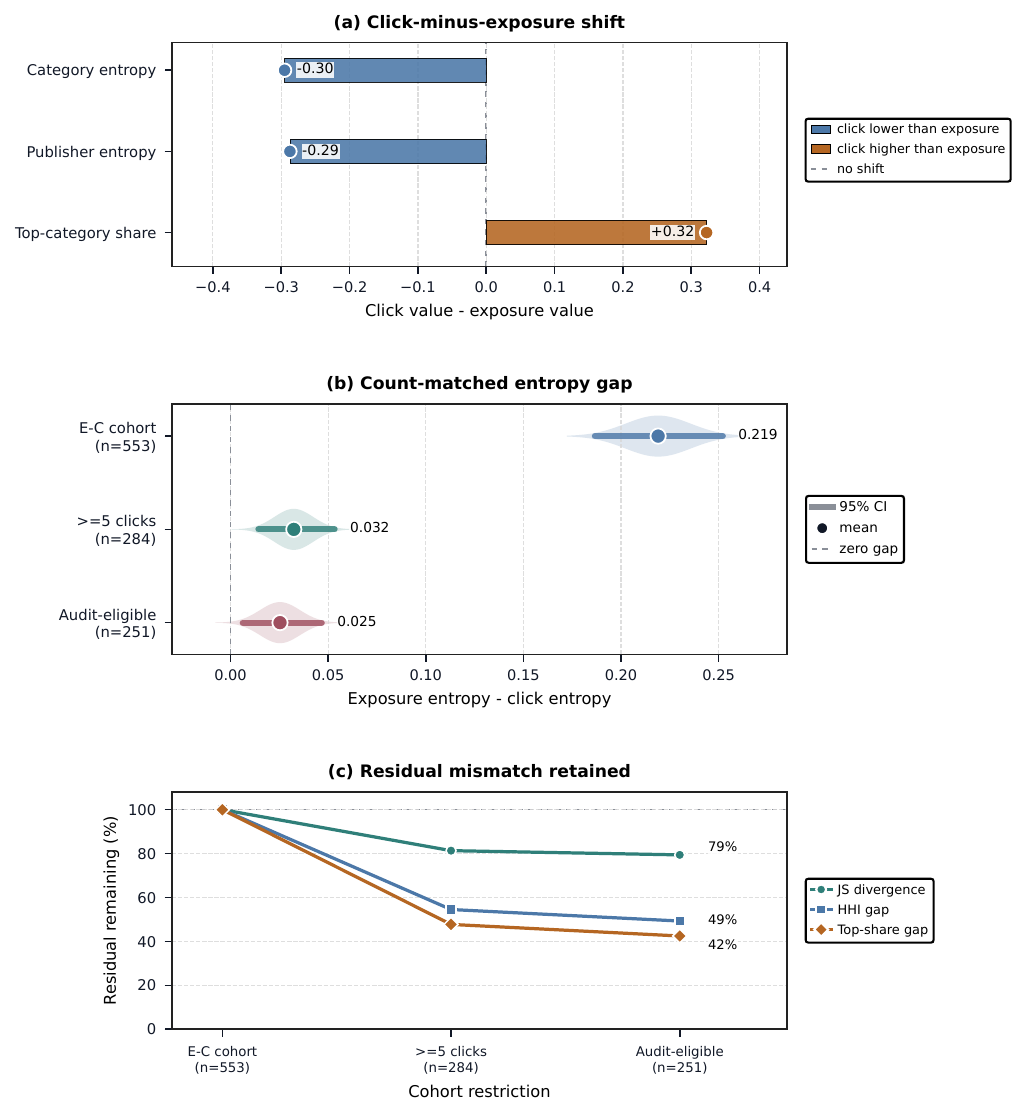}
\caption{Diversity and concentration audit: click-minus-exposure shifts, count-matched entropy gaps, and residual mismatch retained across defined cohort restrictions. Residual remaining in panel (c) is measured relative to the corresponding E-C cohort value set to 100\%; cohort labels report user counts.}
\label{fig:diversity-audit}
\end{figure}

\subsection{RQ4: Divergence Is Associated with Observable User Features}

RQ4 examines whether preference-consumption divergence is associated with observable profile and behavioral features. Each diagnostic task evaluates the same four prespecified model families---logistic regression, Random Forest, Extra Trees, and gradient boosting---and reports the highest mean user-level five-fold cross-validated AUC. Because the high-divergence label is defined from a preference-consumption relationship, the headline same-window AUC is not the claim by itself; the claim is the pattern across label-proximity, dense-history, alternative-label, and future-window checks. Weighted-profile predictors use only distributional properties of the observed profile snapshot, such as preference entropy, maximum preference weight, and salient preference count. Full audit models may include aggregate behavioral summaries, but exclude clicked-item and clicked-category identities, clicked-category weights, preference-click similarity, and other direct alignment quantities used to construct the label. Diagnostic results are reported in Appendix E, while validation settings, model specifications, and feature definitions are documented in Appendices F--G. Within that descriptive setting, the contemporaneous model without weighted-profile features reaches AUC 0.784. Adding observed weighted profile-state features raises same-window audit AUC to 0.859 and AUPRC to 0.861. Weighted profile-state features alone reach AUC 0.696, showing that probability distributions contain more audit signal than basic profile and support-count features alone.

The 0.859 AUC partly reflects the proximity of profile-structure features to the divergence label. Profile structure is informative for preference-consumption divergence, but it is also close to the construct being audited. In addition, the available profile table is a current snapshot rather than a versioned preference history. We therefore treat the model as evidence that divergence is structured, not as a deployable ranking, targeting, or prospective forecasting model.

As an additional bounded validity check, the future-window diagnostic uses January-March behavior and exposure features to identify April-June high-divergence users with AUC 0.731, excluding current weighted-profile snapshot features because historical profile values are unavailable. Because the April-June label still uses the available non-versioned profile snapshot, this is a temporal diagnostic over behavior and exposure features rather than a fully prospective preference-state forecast. A more interpretable top-category mismatch label reaches AUC 0.737, and weighted profile-state features alone reach AUC 0.751 for this simpler label. As an upper-bound diagnostic, an extreme-mismatch task that removes middle cases and compares the top and bottom quartiles of divergence reaches AUC 0.896. This is not a primary modeling result.

\begin{figure}[t]
\centering
\includegraphics[width=\columnwidth,trim=7pt 8pt 7pt 6pt,clip]{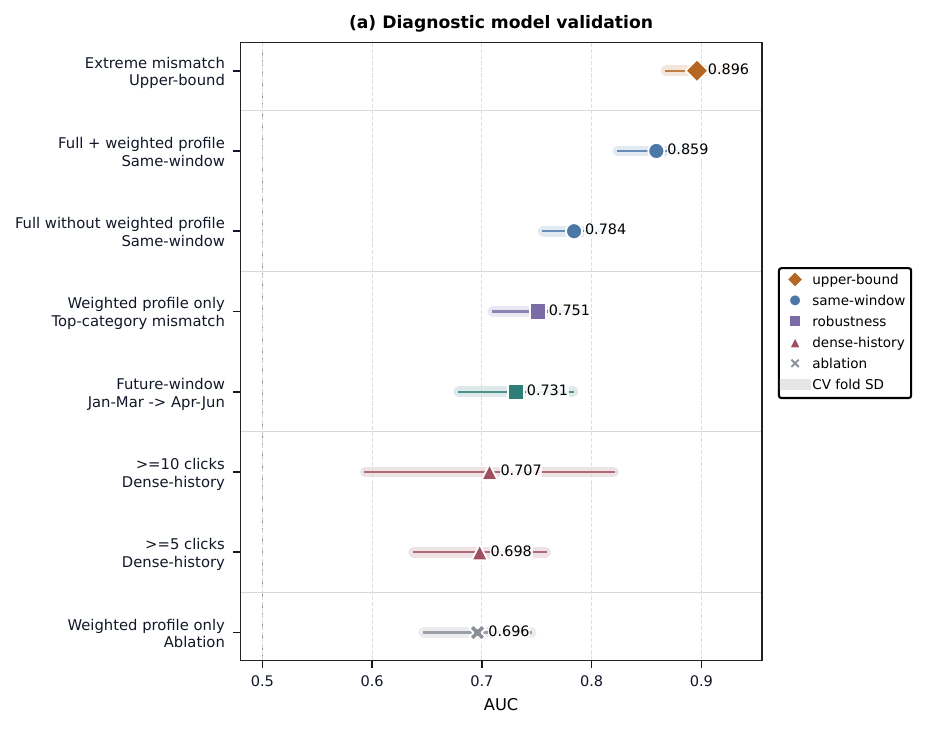}
\caption{Bounded model audit: mean AUC and cross-validation fold standard deviations across upper-bound, same-window, robustness, dense-history, and ablation checks. Each point reports the best mean AUC among the four prespecified model families for that task.}
\label{fig:bounded-signal}
\end{figure}

\newpage
\subsection{User Heterogeneity and Model Stress Tests}

High-divergence users are not simply a more active version of other users. In tertile analysis, the high-divergence segment has mean divergence 0.938 and median click count 2.0, compared with mean divergence 0.604 and median click count 8.0 in the low-divergence segment. Among users with at least five and ten clicks, the weighted full model reaches AUC 0.698 and 0.707, respectively, lower than the full-sample same-window AUC 0.859. The gap is therefore structured, but sparse histories and profile-distribution shape contribute to the easiest classification cases.

\section{Robustness and Sensitivity}

We run robustness checks to address sparsity, metric choice, future-window stability, and activity-volume concerns. The preference-consumption gap is not entirely driven by users with very few clicks: applying the same empty-set rule, top-3 category Jaccard is 0.300 among users with at least five clicks ($n=302$) and 0.310 among those with at least ten ($n=175$). A chance-based permutation check supports the RQ1 interpretation: observed top-category match and mean Jaccard exceed both a shuffled-user preference-set null (61.54\%, 0.161) and a category-frequency, size-preserving null (54.81\%, 0.137). Separately from the pooled entropy values in RQ3, a user-level bootstrap confirms the mean recommendation-click category entropy gap (0.288, 95\% CI [0.254, 0.320]), while count-normalized entropy reduces but does not fully erase the gap among users with at least five clicks. The observed same-day hit rate is 11.29\%, compared with a date-matched logged-list candidate-pool expectation of 9.13\% (95\% CI [8.77\%, 9.53\%]); the paired observed-minus-baseline gap is 2.16 percentage points under user-clustered bootstrap (95\% CI [0.64, 4.37]), indicating signal without exhausting the multi-surface consumption environment. High-divergence classification is not reducible to click volume: removing click\_count, avg\_daily\_clicks, and active\_click\_days from the full audit model yields AUC 0.752. In a separate, less restrictive specification that retains total click count but removes avg\_daily\_clicks and active\_click\_days, AUC is 0.776. These checks reduce the plausibility that the main findings are driven by one threshold, one metric, or raw click volume.

\section{Discussion}

The results support four main conclusions. First, explicit preferences are informative but incomplete: they capture broad orientation better than the breadth and situational variation of actual consumption. The relatively high null expectations show why raw top-category agreement alone can overstate preference-consumption alignment. Second, logged recommendation lists contain clicked articles more often than descriptive candidate-pool and simple popularity baselines predict, yet trace only a modest share of observed clicks. Third, exposure-side diversity should not be conflated with consumption-side diversity: raw entropy gaps shrink under count matching, yet concentration and distributional mismatch remain. Fourth, preference-consumption divergence is associated with observable profile-state and behavioral features, but AUC declines in dense-history and future-window checks. This pattern shows that same-window predictability partly reflects sparse histories and the proximity of profile-state features to the audit label, while some signal remains under stricter validation.

The exposure-consumption diversity gap must therefore be interpreted carefully. Count-matched checks show that sparse click observation explains part of the raw entropy gap, but concentration-based metrics remain positive in the audit-eligible cohort. Diversity is partly a measurement-layer problem and partly a click-allocation pattern.

The PEC gap is therefore not only a behavioral gap; it is also a measurement gap created by the different observability of logged exposure and click consumption. Diversity represented in recommendation logs, observed consumption diversity, and complete attention are different quantities. The paper's contribution is to make those quantities separable rather than collapsing them into a single preference signal.

The contribution is methodological rather than a claim that preference, exposure, and consumption have never previously been distinguished. We do not claim that the observed hit rates, entropy gaps, or AUC values will hold across platforms. Instead, PEC provides a reusable audit protocol for testing whether a platform's own preference, exposure, surface-pathway, and consumption traces can be safely interpreted as aligned. On another platform, the same audit logic could map P to followed topics or declared interests, E to feed impressions, S to homepage, search, or notification pathways, and C to clicks, reads, watch starts, or other observed consumption events.

A key implication is that mobile news is a multi-surface system rather than a single-list system. Headline modules, category tabs, home/feed views, newsroom pages, search, notifications, and article-detail continuations shape what becomes clickable and measurable. Evaluation that ignores surface architecture risks overstating what recommendation logs alone can explain.

Practically, PEC suggests that platforms should avoid using clicks alone as preference proxies, treat recommendation logs as bounded exposure rather than total exposure, specify the measurement layer when making diversity claims, and audit surface-level entry points alongside ranked lists. Design responses should ask where alignment weakens across P, E, S, and C before treating narrow click consumption as user preference or recommendation failure.

\section{Limitations}

This study is observational and cannot establish causal effects of ranking, interface design, or recommendation exposure; the findings are measurement-audit claims about observable traces, not causal estimates of platform influence. A clicked article missing from a same-day recommendation list means only that the item was absent from the logged recommendation list available for that user and date.

Recommendation logs do not capture every possible exposure surface, and list membership does not establish viewport entry or visual inspection. Same-day hit should therefore be read as traceability within available logged recommendation records, not as complete platform-level exposure or conversion.

The candidate-pool and popularity-aware baselines are date-matched diagnostic checks, but they do not fully separate personalized recommendation signal from article-level popularity or editorial salience. Future PEC audits could add editorial and category-matched baselines to further decompose traceability.

The dataset comes from a single mobile news application. Because platform traces reflect both user population and product affordances, the empirical magnitudes may not generalize to platforms with different interfaces, recommendation strategies, or notification systems (Lazer et al. 2021). The reusable claim is PEC separation, not these estimates.

The analysis captures metadata-level category and publisher breadth rather than democratic diversity in the broader normative sense (Helberger 2019). We therefore do not make claims about ideological diversity, stance exposure, or political polarization.

Click logs provide informative but biased and partial evidence about perceived relevance rather than absolute user interest (Joachims et al. 2005). They identify article openings, not dwell time, reading depth, completion, comprehension, or meaningful engagement; we describe the outcome as observed click consumption rather than complete attention.

A limitation of the profile analysis is that the available data contain current profile snapshots rather than complete preference-edit histories. Future instrumentation should log preference-edit events and time-stamped profile snapshots so that initial stated preferences, manual revisions, and observed consumption can be separated more strictly.

Future PEC audits should extend instrumentation in four directions: time-stamped preference-edit histories, impression-level logging across all app surfaces, attention traces such as dwell time and reading depth, and responsibly validated article-level topic, stance, or ideology labels.

The classification model is diagnostic, not operational; its AUC does not support targeting or intervention.

\section{Ethics, Privacy, and Data Availability}

This study analyzes ordinary-use operational logs from a deployed mobile news application as an internal platform audit. The analysis used platform-internal identifiers, reports no direct identifiers, and did not alter users' recommendations, ranking outputs, notifications, or app experience.

No separate research-specific consent was obtained. The logs were collected under internal privacy and data-governance procedures; no intervention or direct participant contact occurred, and only aggregate results are reported.

Data access was limited to authorized platform personnel. Because behavioral news traces remain sensitive after pseudonymization, the paper reports only aggregate statistics and does not disclose individual browsing histories, raw user profiles, device identifiers, or personally identifiable information. The analysis does not attempt to infer protected characteristics such as race, religion, health status, political affiliation, or other sensitive attributes.

Raw logs cannot be publicly released because timestamped news consumption traces may create re-identification risks even after direct identifiers are removed. To support reproducibility, the release package includes preprocessing code, metric definitions, aggregate tables, figure-generation scripts, and synthetic schema examples.

The high-divergence model is framed as an audit instrument, not a targeting system; any future intervention inspired by PEC divergence should be reviewed for transparency, consent expectations, and downstream harms.

\section{Conclusion}

This paper audited a deployed mobile news platform by linking stated preferences, observed weighted profile-state snapshots, logged recommendation exposure, click behavior, and news metadata. Across four analyses, the PEC gap is consistently visible: stated preferences capture dominant interests but not full consumption breadth; logged recommendation exposure carries measurable but partial signal; logged exposure is more diverse than click consumption in the raw logs, while count-adjusted concentration and distributional mismatch remain; and high-divergence users are partly identifiable under bounded diagnostic conditions. The empirical magnitudes are specific to one application and six months of activity, so the contribution is the audit design rather than the point estimates: platforms with preference, exposure, surface, and consumption traces can instantiate the same separation before treating heterogeneous traces as substitutes.

\section*{References}

Abdollahpouri, H.; Burke, R.; and Mobasher, B. 2019. Managing Popularity Bias in Recommender Systems with Personalized Re-Ranking. In Proceedings of the Thirty-Second International Florida Artificial Intelligence Research Society Conference (FLAIRS-32), 413--418. AAAI Press.

An, Z.; Breuhaus, J.; Niu, J.; Sariyuce, A. E.; and Joseph, K. 2024. Curated and Asymmetric Exposure: A Case Study of Partisan Talk during COVID on Twitter. Proceedings of the International AAAI Conference on Web and Social Media, 18(1): 70-85. doi:10.1609/icwsm.v18i1.31298.

Bakshy, E.; Messing, S.; and Adamic, L. A. 2015. Exposure to ideologically diverse news and opinion on Facebook. Science, 348(6239): 1130--1132. doi:10.1126/science.aaa1160.

Bandy, J.; and Diakopoulos, N. 2020. Auditing News Curation Systems: A Case Study Examining Algorithmic and Editorial Logic in Apple News. Proceedings of the International AAAI Conference on Web and Social Media, 14(1): 36--47. doi:10.1609/icwsm.v14i1.7277.

Bandy, J.; and Lazovich, T. 2023. Exposure to Marginally Abusive Content on Twitter. Proceedings of the International AAAI Conference on Web and Social Media, 17(1): 24-33. doi:10.1609/icwsm.v17i1.22123.

Breiman, L. 2001. Random forests. Machine Learning, 45(1): 5-32. doi:10.1023/A:1010933404324.

DeVito, M. A. 2017. From editors to algorithms: A values-based approach to understanding story selection in the Facebook News Feed. Digital Journalism, 5(6): 753-773. doi:10.1080/21670811.2016.1178592.

Diakopoulos, N. 2015. Algorithmic accountability: Journalistic investigation of computational power structures. Digital Journalism, 3(3): 398-415. doi:10.1080/21670811.2014.976411.

Duskin, K.; Schafer, J. S.; Efstratiou, A.; West, J. D.; and Spiro, E. S. 2026. The Role of Follow Networks and Twitter's Content Recommender on Partisan Skew and Rumor Exposure During the 2022 U.S. Midterm Election. Proceedings of the International AAAI Conference on Web and Social Media, 20(1): 658-676. doi:10.1609/icwsm.v20i1.42659.

Flaxman, S.; Goel, S.; and Rao, J. M. 2016. Filter bubbles, echo chambers, and online news consumption. Public Opinion Quarterly, 80(S1): 298-320. doi:10.1093/poq/nfw006.

Geurts, P.; Ernst, D.; and Wehenkel, L. 2006. Extremely randomized trees. Machine Learning, 63(1): 3-42. doi:10.1007/s10994-006-6226-1.

Guess, A. M.; Malhotra, N.; Pan, J.; Barberá, P.; Allcott, H.; Brown, T.; Crespo-Tenorio, A.; Dimmery, D.; Freelon, D.; Gentzkow, M.; et al. 2023. How do social media feed algorithms affect attitudes and behavior in an election campaign? Science, 381(6656): 398-404. doi:10.1126/science.ade7138.

Hart, W.; Albarracín, D.; Eagly, A. H.; Brechan, I.; Lindberg, M. J.; and Merrill, L. 2009. Feeling validated versus being correct: A meta-analysis of selective exposure to information. Psychological Bulletin, 135(4): 555-588. doi:10.1037/a0015701.

Helberger, N. 2019. On the democratic role of news recommenders. Digital Journalism, 7(8): 993-1012. doi:10.1080/21670811.2019.1623700.

Hosseinmardi, H.; Ghasemian, A.; Rivera-Lanas, M.; Horta Ribeiro, M.; West, R.; and Watts, D. J. 2024. Causally estimating the effect of YouTube's recommender system using counterfactual bots. Proceedings of the National Academy of Sciences, 121(8): e2313377121. doi:10.1073/pnas.2313377121.

Huszár, F.; Ktena, S. I.; O'Brien, C.; Belli, L.; Schlaikjer, A.; and Hardt, M. 2022. Algorithmic amplification of politics on Twitter. Proceedings of the National Academy of Sciences, 119(1): e2025334119. doi:10.1073/pnas.2025334119.

Jannach, D.; and Adomavicius, G. 2016. Recommendations with a purpose. In Proceedings of the ACM Conference on Recommender Systems (RecSys), 7-10. doi:10.1145/2959100.2959186.

Joachims, T.; Granka, L.; Pan, B.; Hembrooke, H.; and Gay, G. 2005. Accurately interpreting clickthrough data as implicit feedback. In Proceedings of the ACM SIGIR Conference on Research and Development in Information Retrieval, 154-161. doi:10.1145/1076034.1076063.

Knijnenburg, B. P.; Willemsen, M. C.; Gantner, Z.; Soncu, H.; and Newell, C. 2012. Explaining the user experience of recommender systems. User Modeling and User-Adapted Interaction, 22(4-5): 441-504. doi:10.1007/s11257-011-9118-4.

Lazer, D. M. J.; Hargittai, E.; Freelon, D.; González-Bailón, S.; Munger, K.; Ognyanova, K.; and Radford, J. 2021. Meaningful measures of human society in the twenty-first century. Nature, 595(7866): 189-196. doi:10.1038/s41586-021-03660-7.

Lin, J. 1991. Divergence measures based on the Shannon entropy. IEEE Transactions on Information Theory, 37(1): 145-151. doi:10.1109/18.61115.

Milli, S.; Carroll, M.; Wang, Y.; Pandey, S.; Zhao, S.; and Dragan, A. D. 2025. Engagement, user satisfaction, and the amplification of divisive content on social media. PNAS Nexus, 4(3): pgaf062. doi:10.1093/pnasnexus/pgaf062.

Möller, J.; Trilling, D.; Helberger, N.; and van Es, B. 2018. Do not blame it on the algorithm: An empirical assessment of multiple recommender systems and their impact on content diversity. Information, Communication \& Society, 21(7): 959-977. doi:10.1080/1369118X.2018.1444076.

Robertson, R. E.; Green, J.; Ruck, D. J.; Ognyanova, K.; Wilson, C.; and Lazer, D. 2023. Users choose to engage with more partisan news than they are exposed to on Google Search. Nature, 618(7964): 342-348. doi:10.1038/s41586-023-06078-5.

Sandvig, C.; Hamilton, K.; Karahalios, K.; and Langbort, C. 2014. Auditing algorithms: Research methods for detecting discrimination on internet platforms. Paper presented at the Data and Discrimination preconference, 64th Annual Meeting of the International Communication Association, Seattle, WA, May 22.

Schlessinger, J.; Garimella, K.; Jakesch, M.; and Eckles, D. 2023. Effects of Algorithmic Trend Promotion: Evidence from Coordinated Campaigns in Twitter's Trending Topics. Proceedings of the International AAAI Conference on Web and Social Media, 17(1): 777-786. doi:10.1609/icwsm.v17i1.22187.

Vrijenhoek, S.; Kaya, M.; Metoui, N.; Möller, J.; Odijk, D.; and Helberger, N. 2021. Recommenders with a mission: Assessing diversity in news recommendations. In Proceedings of the ACM SIGIR Conference on Human Information Interaction and Retrieval (CHIIR), 173-183. doi:10.1145/3406522.3446019.

\clearpage
\section*{Checklist}

1. For most authors...

(a) Would answering this research question advance science without violating social contracts, such as privacy norms, unfair profiling, socio-economic divides, or disrespect to societies or cultures? Yes. The study is an internal platform audit using platform-internal identifiers and aggregate reporting; no direct identifiers are reported. It does not infer protected attributes or manipulate the user experience.

(b) Do your main claims in the abstract and introduction accurately reflect the paper's contributions and scope? Yes. The paper claims a measurement audit framework and bounded empirical findings, not causal recommendation effects, ideology, or production targeting.

(c) Do you clarify how the proposed methodological approach is appropriate for the claims made? Yes. Sections 3-5 define the PEC protocol, logged-exposure scope, surface pathways, metric construction, baselines, and bounded modeling interpretation.

(d) Do you clarify possible artifacts in the data, given population-specific distributions? Yes. The paper discusses single-platform scope, partial recommendation user-days, surface observability, click sparsity, and current-profile snapshot limitations.

(e) Did you describe the limitations of your work? Yes. Section 8 and Table~\ref{tab:claim-map} state that logged exposure is not total exposure, clicks are incomplete attention proxies, and category/publisher diversity is not ideology.

(f) Did you discuss any potential negative societal impacts of your work? Yes. The ethics section discusses risks of using divergence outputs for targeting and emphasizes transparency, consent expectations, and user-protection safeguards.

(g) Did you discuss any potential misuse of your work? Yes. The paper states that the diagnostic model should not be used for targeting or intervention without additional review.

(h) Did you describe steps taken to prevent or mitigate potential negative outcomes, such as documentation, anonymization, responsible release, access control, and reproducibility? Yes. Raw logs are not released; the reproducibility plan uses preprocessing code, aggregate tables, figure scripts, release-safe notebook walkthroughs, and synthetic schemas.

(i) Have you read the ethics review guidelines and ensured that your paper conforms to them? Yes. The study is framed as a non-interventional audit with aggregate reporting and explicit data-governance constraints.

2. Additionally, if your study involves hypotheses testing...

(a) Did you clearly state the assumptions underlying all theoretical results? NA for theoretical results. The paper presents an empirical measurement audit with stated observability assumptions. For the empirical permutation tests, the null constructions, preserved quantities, number of permutations, confidence intervals, and one-sided empirical p-values are specified in Appendix F.

(b) Have you provided justifications for all theoretical results? NA. No theoretical results are claimed. The empirical quantities are justified through metric definitions, permutation/null checks, robustness checks, and explicit scope conditions.

(c) Did you discuss competing hypotheses or theories that might challenge or complement your theoretical results? NA for theoretical results. The paper nevertheless discusses competing explanations for the observed gaps, including sparse click histories, partial exposure logging, surface-pathway effects, and current-profile snapshot ambiguity.

(d) Have you considered alternative mechanisms or explanations that might account for the same outcomes observed in your study? Yes. The paper treats alternative mechanisms as central validity concerns and evaluates them through count matching, candidate-pool and popularity-aware baselines, click-threshold cohorts, alternative labels, and future-window checks.

(e) Did you address potential biases or limitations in your theoretical framework? Yes. The framework is explicitly bounded by single-platform scope, incomplete exposure observability, click-as-attention limitations, and metadata-level rather than ideology-level diversity.

(f) Have you related your theoretical results to the existing literature in social science? NA for theoretical results. The measurement framework is related to prior social computing, selective exposure, platform audit, recommender evaluation, and digital trace validity literature.

(g) Did you discuss the implications of your theoretical results for policy, practice, or further research in the social science domain? NA for theoretical results. The discussion explains practical implications for platform audit design, interface-aware evaluation, and future longitudinal instrumentation.

3. Additionally, if you are including theoretical proofs...

(a) Did you state the full set of assumptions of all theoretical results? NA. The paper does not include formal theoretical results.

(b) Did you include complete proofs of all theoretical results? NA. No formal theoretical proofs are included.

4. Additionally, if you ran machine learning experiments...

(a) Did you include code, data, and instructions needed to reproduce the main experimental results? Partially. Raw operational logs cannot be released because timestamped user traces create re-identification risk, but the public package provides metric code, aggregate tables, figure-generation scripts, release-safe notebooks, and synthetic schema examples. Authorized reruns additionally use output-stripped notebooks and processed private logs.

(b) Did you specify all training details, data splits, hyperparameters, and how they were chosen? Yes. Detailed results are reported in Appendix E, with training and feature specifications in Appendices F--G. Broader sensitivity and stress-test outputs are retained in the project tables.

(c) Did you report error bars or uncertainty? Yes. The paper reports user-level bootstrap intervals for diversity quantities, descriptive interval estimates for traceability checks, and robustness results across cohorts, labels, ablations, and future-window checks.

(d) Did you include the total amount of compute and type of resources used? Yes. A full measured rerun, including 10,000 RQ1 permutations and all diagnostic checks, took approximately 20 minutes on a 12-core Apple M2 Pro workstation with 16 GB RAM. The pipeline uses CPU-based tabular aggregation and classification with no GPU training or large-scale model pretraining.

(e) Do you justify how the proposed evaluation is sufficient and appropriate to the claims made? Yes. The evaluation is explicitly diagnostic: candidate-pool and popularity-aware baselines test traceability, count matching tests observation-count confounding, and future-window checks bound model claims.

(f) Do you discuss the cost of misclassification and fault tolerance? Yes. Because the model is not a production targeting model, misclassification cost is discussed as a reason to keep the result diagnostic and non-operational.

5. Additionally, if you are using existing assets or curating/releasing new assets, without compromising anonymity...

(a) If your work uses existing assets, did you cite the creators? Yes. The paper cites prior studies, recommender-system evaluation work, platform-audit research, exposure-measurement studies, and model families used in the analysis. The operational platform logs are internal first-party records rather than an externally created public dataset.

(b) Did you mention the license of the assets? Yes. The release code is provided under the MIT License; release-safe aggregate tables and synthetic schema examples are provided under CC BY 4.0. These licenses do not apply to raw operational logs, which remain non-public under platform data-governance constraints. The paper does not redistribute third-party datasets or pretrained model assets.

(c) Did you include any new assets in the supplemental material or as a URL? Partially. The reproducibility package includes preprocessing scripts, metric scripts, release-safe notebook walkthroughs, aggregate tables used for manuscript figures, and synthetic schemas. Authorized private rerun materials use output-stripped notebooks with approved processed logs. The public package does not include timestamped raw user logs or identifiable behavioral traces.

(d) Did you discuss whether and how consent was obtained from people whose data you are using or curating? Yes. No separate research-specific consent was obtained. The analysis used operational logs collected during ordinary service use under applicable privacy and data-governance procedures. The study involved no intervention or direct participant contact, and only aggregate results are reported.

(e) Did you discuss whether the data you are using or curating contains personally identifiable information or offensive content? Yes. The paper states that timestamped user traces can create re-identification risk even after pseudonymization; therefore raw logs are not released. The news metadata may include article text fields, but the analysis reports category, publisher, recommendation, surface, and aggregate behavior metrics rather than quoting private user content.

(f) If you are curating or releasing new datasets, did you discuss how you intend to make your datasets FAIR? Partially. A public user-level dataset is not released because of privacy and platform-governance constraints. To support inspection and reuse without exposing raw logs, the project documents metric definitions, processing code, aggregate tables, and synthetic schemas that mirror the table structure.

(g) If you are curating or releasing new datasets, did you create a Datasheet for the Dataset? NA for raw dataset release. Because the paper does not release a raw dataset, a full datasheet for public user-level data is not applicable. The appendix instead provides schema, preprocessing, observability, and release-boundary documentation for the reproducibility artifacts.

6. Additionally, if you used crowdsourcing or conducted research with human subjects, without compromising anonymity...

(a) Did you include the full text of instructions given to participants and screenshots? NA. The study does not use crowdsourcing, experimental tasks, surveys, or newly issued participant instructions. It audits existing platform logs collected during ordinary app use.

(b) Did you describe any potential participant risks, with mentions of Institutional Review Board approvals? NA. This retrospective observational study analyzed pseudonymized operational logs under authorized internal access. It involved no participant recruitment, intervention, or direct contact, and results are reported only in aggregate.

(c) Did you include the estimated hourly wage paid to participants and the total amount spent on participant compensation? NA. No participants were recruited or compensated.

(d) Did you discuss how data is stored, shared, and deidentified? Yes. The paper states that raw logs are not publicly released; the reproducibility plan uses aggregate tables, synthetic schemas, and executable code. Access to raw operational logs is limited to authorized platform personnel for research and auditing purposes.

\appendix
\renewcommand{\thesection}{Appendix \Alph{section}:}
\section{Log Schema and Preprocessing}

The appendix documents implementation details for reproducibility and validity checking. It does not introduce additional substantive claims beyond the main paper.

Trace construction begins by separating the logs into user profile records, item-level recommendation records, click or behavior events, and article metadata. User profile records provide explicit category and source preferences plus observed weighted profile-state distributions. Recommendation records are exploded from ranked user-date lists into item-level rows with user identifier, recommendation date, rank, and news identifier. Click events are deduplicated and then joined to article metadata using news identifiers. Article metadata supply category, publisher, publication date, language, summary, tag, breaking-news flag, and available text fields.

Deduplication uses timestamp, user identifier, session identifier, event name, news identifier, and date keys when available. Comparable click rows for RQ2 require a clicked article with metadata and a same-day recommendation list for that user. Missing or invalid surface metadata are retained as other/unknown rather than silently dropped.

\section{Surface Assignment Rules}

Surface assignment uses the final recorded click-time metadata and assigns exactly one pathway per matched click. The priority rule uses notification type, source-path metadata, section labels, and headline-related fields to avoid double-counting a click across surfaces. Surface labels should therefore be interpreted as observed entry pathways, not as full impression histories.

The newsroom label captures in-app clicks whose metadata indicate the newsroom tab or newsroom-related sections. In the current logs, this surface is identified by internal newsroom-path metadata and multiple newsroom-section values, which are abstracted here to avoid exposing product-specific identifiers. We therefore interpret this surface as an in-app newsroom pathway, not as external inbound traffic or an outbound click to another news site.

\section{Metric Definitions}

The top-clicked-category set contains each user's top three clicked categories, or fewer when fewer clicked categories are observed. Ties are handled deterministically by frequency and stable category ordering so that repeated runs produce the same set. The 19 affected records contain a present but empty explicit-selection field and are marked onboarding-complete; they are therefore treated as observed empty current states rather than missing profile rows. Because each clicked-category set is nonempty, their set overlap is well defined as $J_u=0$. The same rule is used in the 698-user RQ1 mean and the RQ4 divergence label ($D_u=1-J_u$); the 679-user nonempty-set RQ1 estimate is reported as a sensitivity check. We do not infer from the current snapshot whether an empty state reflects later clearing or a legacy state. User-level exposure distributions are computed over the user's logged recommendation items across the observation window, while click distributions are computed over matched clicked articles.

Entropy is normalized by the logarithm of the number of nonzero categories or publishers in the corresponding sampled distribution; one-label distributions are assigned entropy 0. HHI is the sum of squared shares, so larger values indicate greater concentration; top-share is the largest category share. Jensen-Shannon divergence is computed with equal weights and base-2 logarithms over the union of observed categories in the user's exposure and click distributions. Because the mixture distribution assigns positive mass wherever either input distribution has positive mass, JS divergence is finite without additive smoothing. We report divergence itself, not its square-root distance. We therefore interpret entropy as normalized within-distribution evenness, while HHI, top-share, effective categories, and JS divergence carry the main support-sensitive concentration and mismatch checks.

Count matching repeatedly samples each user's logged exposure items down to that user's number of observed clicks and recomputes the target metric. This check is used to distinguish diversity gaps caused by sparse click observation from gaps that remain after exposure and consumption are placed on comparable count scales.

\section{Traceability Baselines}

All traceability baselines use the same comparable-click rows as RQ2 and preserve each user-date recommendation list size. The date-matched logged-list candidate-pool baseline computes the expected probability that a clicked article would appear in a list of the same size drawn from the same-date recommendation candidate pool. The candidate pool is the union of articles appearing in any logged recommendation list on that date after comparable-click metadata filters; it is not the full news corpus. The release-safe baseline does not reconstruct every production eligibility rule or language/market serving constraint, so it should be read as a logged-list candidate-pool check rather than a full serving simulator. If a clicked article is not in that date's candidate pool, its baseline probability is zero.

Popularity-aware baselines replace user-specific recommendation output with non-personalized popularity signals while preserving the user-date list size. The leave-user-out same-day recommendation-popularity baseline ranks articles by how often they appear in other users' recommendation lists on the same date. The prior-day click-popularity baseline ranks articles by click frequency on the previous day. The prior-week click-popularity baseline ranks articles by click frequency over the prior seven days. These checks are descriptive traceability comparisons rather than causal estimates of personalization effects.

\section{Supplementary Result Tables}

Tables in this section collect the detailed numerical summaries referenced in the main text. They are grouped by claim: preference alignment, exposure traceability, diversity checks, surface pathways, and diagnostic models.

\begin{table}[!h]
\centering
\caption{Preference-consumption alignment.}
\label{tab:preference-alignment}
\scriptsize
\setlength{\tabcolsep}{4pt}
\renewcommand{\arraystretch}{1.12}
\begin{tabular}{@{}>{\raggedright\arraybackslash}p{0.68\columnwidth}>{\raggedright\arraybackslash}p{0.22\columnwidth}@{}}
\toprule
\textbf{Metric} & \textbf{Result} \\
\midrule
Users with profile and matched clicks & 698 \\
Top consumed category in stated preferences & 77.22\% \\
Users in primary mean Jaccard & 698 \\
Empty stated-category sets ($J_u=0$) & 19 \\
Mean category Jaccard overlap & 0.2283 \\
Nonempty-set sensitivity ($n=679$) & 0.2347 \\
Median category Jaccard overlap & 0.2000 \\
Category weight: top clicked / top-three mass & 0.143 / 0.264 \\
Category cosine / JS divergence & 0.452 / 0.579 \\
Publisher top-click weight / cosine & 0.014 / 0.080 \\
\bottomrule
\end{tabular}
\tablenote{Primary Jaccard includes all 698 users and assigns $J_u=0$ to the 19 observed empty stated-category states. The nonempty sensitivity restricts to users with at least one stated category.}
\end{table}

\begin{table*}[t]
\centering
\caption{Claim map for the four PEC audits.}
\label{tab:claim-map}
\scriptsize
\setlength{\tabcolsep}{4pt}
\renewcommand{\arraystretch}{1.12}
\begin{tabular}{@{}>{\raggedright\arraybackslash}p{0.19\textwidth}>{\raggedright\arraybackslash}p{0.51\textwidth}>{\raggedright\arraybackslash}p{0.21\textwidth}@{}}
\toprule
\textbf{Audit claim} & \textbf{Main evidence} & \textbf{Scope condition} \\
\midrule
Stated preferences recover dominant interests, but miss broader click variation. & Top-category match is 77.22\%; mean Jaccard is 0.2283 ($n=698$; nonempty sensitivity 0.2347); category cosine is 0.452; weighted-profile-only AUC is 0.696. & Current profile snapshots, not edit histories or time-stamped preference changes. \\
Logged exposure traces clicks only partially. & Same-day hit is 11.29\% versus a 9.13\% candidate-pool expectation and simple popularity baselines; top-5 lift is 1.47x; matched clicks arrive mainly through headline, category, and home/feed pathways. & Logged exposure is not total exposure; no causal attribution or editorial-salience decomposition. \\
Exposure appears more diverse than clicks, but the robust claim is count-adjusted mismatch. & Pooled entropy is 0.835 for exposure versus 0.540 for clicks; after count matching, HHI gap 0.082, top-share gap 0.069, and JS 0.471 remain. & Cohort-dependent audit result; category/publisher gaps are not ideology or complete attention. \\
High preference-consumption divergence is partly predictable. & Full model with weighted profile reaches AUC 0.859; future-window diagnostic reaches 0.731; click-threshold checks reach 0.698-0.707. & Diagnostic audit signal, not production targeting, ranking, or causal explanation. \\
\bottomrule
\end{tabular}
\tablenote[0.93\textwidth]{Each row states the audit claim, the numerical evidence used to support it, and the boundary that limits interpretation. The scope column marks what should not be inferred from the evidence; these are measurement-audit claims, not causal effect estimates.}
\vspace{-0.5em}
\end{table*}

\begin{table}[!htbp]
\centering
\caption{Exposure-consumption traceability.}
\scriptsize
\setlength{\tabcolsep}{4pt}
\renewcommand{\arraystretch}{1.12}
\begin{tabular}{@{}>{\raggedright\arraybackslash}p{0.68\columnwidth}>{\raggedright\arraybackslash}p{0.22\columnwidth}@{}}
\toprule
\textbf{Metric} & \textbf{Result} \\
\midrule
Matched clicks & 17,232 \\
Clicks with same-day recommendation lists & 11,440 \\
Same-day recommendation hit rate (comparable clicks; $n=11{,}440$) & 11.29\% \\
Six-month recommendation-history overlap (all matched clicks; $n=17{,}232$) & 14.37\% \\
\bottomrule
\end{tabular}
\tablenote{Same-day hit is computed only for clicks with a same-day logged recommendation list. The six-month overlap relaxes same-day timing and therefore uses all matched clicks.}
\end{table}

\begin{table}[!htbp]
\centering
\caption{Top-k recommendation traceability.}
\scriptsize
\setlength{\tabcolsep}{4pt}
\renewcommand{\arraystretch}{1.12}
\begin{tabular}{@{}>{\raggedright\arraybackslash}p{0.22\columnwidth}>{\raggedright\arraybackslash}p{0.22\columnwidth}>{\raggedright\arraybackslash}p{0.24\columnwidth}>{\raggedright\arraybackslash}p{0.18\columnwidth}@{}}
\toprule
\textbf{Top-k} & \textbf{Hit} & \textbf{Candidate-pool} & \textbf{Lift} \\
\midrule
3 & 3.28\% & 2.27\% & 1.45x \\
5 & 4.99\% & 3.41\% & 1.47x \\
10 & 7.76\% & 5.60\% & 1.39x \\
20 & 11.05\% & 8.66\% & 1.28x \\
\bottomrule
\end{tabular}
\tablenote{Candidate-pool values are date-matched expectations for same-size lists drawn from logged recommendation candidates. Lift is observed hit divided by this expectation, not causal attribution.}
\end{table}

\begin{table}[!htbp]
\centering
\caption{Count-matched diversity checks.}
\scriptsize
\setlength{\tabcolsep}{2pt}
\renewcommand{\arraystretch}{1.12}
\begin{tabular}{@{}>{\raggedright\arraybackslash}p{0.24\columnwidth}>{\raggedright\arraybackslash}p{0.10\columnwidth}>{\raggedright\arraybackslash}p{0.14\columnwidth}>{\raggedright\arraybackslash}p{0.12\columnwidth}>{\raggedright\arraybackslash}p{0.17\columnwidth}>{\raggedright\arraybackslash}p{0.10\columnwidth}@{}}
\toprule
\textbf{Cohort} & \textbf{Users} & \textbf{Entropy} & \textbf{HHI} & \textbf{Top-share} & \textbf{JS div.} \\
\midrule
Click $\geq$5 & 284 & 0.032 & 0.091 & 0.077 & 0.483 \\
Audit-eligible & 251 & 0.025 & 0.082 & 0.069 & 0.471 \\
\bottomrule
\end{tabular}
\tablenote{Entropy is exposure minus click entropy after count matching. HHI and top-share are click concentration minus matched-exposure concentration, so positive values mean clicks are more concentrated; JS compares matched exposure and click distributions.}
\end{table}

\begin{table}[!htbp]
\centering
\caption{Click pathways by app surface.}
\scriptsize
\setlength{\tabcolsep}{2pt}
\renewcommand{\arraystretch}{1.12}
\begin{tabular}{@{}>{\raggedright\arraybackslash}p{0.21\columnwidth}>{\raggedright\arraybackslash}p{0.14\columnwidth}>{\raggedright\arraybackslash}p{0.13\columnwidth}>{\raggedright\arraybackslash}p{0.17\columnwidth}>{\raggedright\arraybackslash}p{0.18\columnwidth}@{}}
\toprule
\textbf{Surface} & \textbf{Clicks} & \textbf{Share} & \textbf{Same-day hit} & \textbf{Pref-click cosine} \\
\midrule
Headline & 11,270 & 65.4\% & 10.05\% & 0.424 \\
Category & 2,421 & 14.0\% & 17.93\% & 0.474 \\
Home/feed & 1,611 & 9.3\% & 12.19\% & 0.484 \\
Newsroom & 1,203 & 7.0\% & 9.05\% & 0.445 \\
Article detail & 712 & 4.1\% & 8.10\% & 0.473 \\
\bottomrule
\end{tabular}
\tablenote{Clicks and shares describe matched clicks assigned to each surface. Same-day hit uses comparable surface clicks; preference-click cosine uses user-surface groups with at least two clicks. The five displayed surfaces cover the major observed pathways.}
\end{table}

\begin{table}[!htbp]
\centering
\caption{Weighted profile model checks.}
\scriptsize
\setlength{\tabcolsep}{2pt}
\renewcommand{\arraystretch}{1.12}
\begin{tabular}{@{}>{\raggedright\arraybackslash}p{0.50\columnwidth}>{\raggedright\arraybackslash}p{0.11\columnwidth}>{\raggedright\arraybackslash}p{0.11\columnwidth}>{\raggedright\arraybackslash}p{0.13\columnwidth}@{}}
\toprule
\textbf{Model check} & \textbf{Users} & \textbf{AUC} & \textbf{AUPRC} \\
\midrule
Basic profile + support counts & 698 & 0.544 & 0.526 \\
Weighted profile-state only & 698 & 0.696 & 0.678 \\
Basic profile + weighted profile-state & 698 & 0.698 & 0.683 \\
Profile + click count + weighted & 698 & 0.823 & 0.837 \\
Full audit + weighted profile-state & 698 & 0.859 & 0.861 \\
Full audit + weighted, clicks $\geq$5 & 302 & 0.698 & 0.518 \\
Full audit + weighted, clicks $\geq$10 & 175 & 0.707 & 0.535 \\
\bottomrule
\end{tabular}
\tablenote{Rows report the best mean AUC among four prespecified model families. Selected models are Logistic for the first three rows, Random Forest for the next three, and gradient boosting for clicks $\geq$10. Dense-history rows are not directly comparable to the full 698-user task.}
\end{table}

\begin{table}[!htbp]
\centering
\caption{Alternative audit labels.}
\scriptsize
\setlength{\tabcolsep}{2pt}
\renewcommand{\arraystretch}{1.12}
\begin{tabular}{@{}>{\raggedright\arraybackslash}p{0.32\columnwidth}>{\raggedright\arraybackslash}p{0.10\columnwidth}>{\raggedright\arraybackslash}p{0.20\columnwidth}>{\raggedright\arraybackslash}p{0.10\columnwidth}>{\raggedright\arraybackslash}p{0.12\columnwidth}@{}}
\toprule
\textbf{Audit task} & \textbf{Users} & \textbf{Model} & \textbf{AUC} & \textbf{AUPRC} \\
\midrule
Future-window high-divergence & 320 & extra trees & 0.731 & 0.739 \\
Top-category mismatch & 124 & extra trees & 0.737 & 0.598 \\
Weighted profile only $\rightarrow$ top-category mismatch & 698 & Logistic & 0.751 & 0.509 \\
Extreme mismatch / upper-bound & 372 & random forest & 0.896 & 0.903 \\
\bottomrule
\end{tabular}
\tablenote{Extreme mismatch is an upper-bound quartile-tail diagnostic with retained threshold ties (372 users), not a main performance claim or a direct comparison to the future-window task.}
\end{table}

\FloatBarrier
\onecolumn
\setlength{\multicolsep}{0pt}
\begin{multicols}{2}
\raggedcolumns
\section{Robustness Checks}

The robustness suite tests whether the PEC findings are artifacts of sparse clicks, metric choice, baseline choice, future window, or modeling label. These checks support the paper's bounded interpretation rather than adding new headline claims.

\par\vspace{0.7em}
\noindent
\begin{minipage}{\columnwidth}
\centering
\captionof{table}{Robustness and sensitivity checks.}
\scriptsize
\setlength{\tabcolsep}{3pt}
\renewcommand{\arraystretch}{1.12}
\begin{tabular}{@{}>{\raggedright\arraybackslash}p{0.20\columnwidth}>{\raggedright\arraybackslash}p{0.47\columnwidth}>{\raggedright\arraybackslash}p{0.24\columnwidth}@{}}
\toprule
\textbf{Check} & \textbf{Purpose} & \textbf{Primary artifact} \\
\midrule
Click thresholds & Test whether preference-consumption gaps are driven by very sparse users. & click-threshold and dense-history tables \\
Count matching & Compare exposure and click diversity on equal observation counts. & aggregate diversity summary \\
Baselines & Separate logged recommendation traceability from simple candidate-pool and popularity effects. & baseline comparison table \\
Future-window validation & Bound the diagnostic model with a forward-looking label window. & model validation table \\
Feature ablation & Check whether AUC is reducible to click volume. & feature-ablation summary \\
\bottomrule
\end{tabular}
\end{minipage}
\tablenote{This table summarizes robustness families; the numerical results appear in Appendix E, and the implementation settings are described below.}
\par\vspace{0.8em}

\textbf{Modeling and null-check details.}

All reported diagnostic tasks use user-level five-fold stratified cross-validation and evaluate the same four prespecified model families: balanced logistic regression, Random Forest, Extra Trees, and gradient boosting. Candidate specifications and seeds are fixed before cross-validation; the reported task-level result is the candidate with the highest mean cross-validated AUC. The selected models are Random Forest for unweighted and weighted same-window, clicks $\geq$5, and extreme mismatch; gradient boosting for clicks $\geq$10; Extra Trees for future-window and top-category mismatch; and logistic regression for the weighted-profile-only checks. Random Forest uses 400--600 trees, minimum leaf size 4--6, and balanced-subsample weights across the specified pipelines. Extra Trees uses 700 trees, minimum leaf size 3, \texttt{max\_features="sqrt"}, balanced weights, \texttt{bootstrap=False}, and seed 92. Seeds are used for reproducibility, not tuning; because model selection and evaluation use the same cross-validation summaries, these AUCs are descriptive diagnostic estimates rather than unbiased post-selection performance estimates. RQ1 null checks use 10,000 fixed-seed permutations. The shuffled-set null permutes complete stated-category sets across users while holding click outcomes fixed; the category-frequency null samples same-size stated-category sets from the empirical marginal preference distribution. Empty stated sets contribute zero overlap in the observed and null estimates. Count matching uses 300 fixed-seed without-replacement draws per eligible user; diversity CIs use 2,000 user bootstrap resamples. The RQ2 paired traceability check uses 5,000 user-clustered bootstrap resamples over comparable click rows. For top-category match, null means are 61.54\% [58.88, 64.04] and 54.81\% [52.01, 57.74], below the observed 77.22\% ($p < 0.001$). Mean-Jaccard null means are 0.161 [0.153, 0.169] and 0.137 [0.129, 0.145], below the observed 0.2283 ($p < 0.001$); excluding empty stated sets yields observed mean 0.2347.

\section{Model Features and Labels}

The high-divergence label is $D_u=1-J_u$ with a median split for the main same-window task. Users strictly above the full-cohort median are labeled positive, and median ties are assigned negative. Empty stated-category sets are assigned $J_u=0$ and $D_u=1$ for this RQ4 label. Positive-class prevalence is 45.6\% for the main same-window task, 46.3\% for the future-window task, 25.5\% and 19.4\% for the click $\geq$5 and click $\geq$10 dense-history checks, and 22.8\% for the top-category-mismatch task. The extreme-mismatch diagnostic compares users at or below the 25th percentile and at or above the 75th percentile of divergence; ties at the discrete quartile thresholds are retained, yielding a near-balanced 372-user task. The future-window task predicts April-June labels from January-March behavior and exposure features.

Full audit features include basic profile indicators and category/source profile-support counts; weighted-profile summaries such as entropy, maximum weight, and salient-count measures; logged-exposure counts and diversity summaries; click volume, active-day counts, click timing, and surface-use summaries. The two support counts use the larger observed support across the explicit-selection and weighted-profile fields; they are not the explicit stated-set cardinality used to construct Jaccard. These aggregate behavioral summaries are retained, but clicked-item identities, clicked-category identities, clicked-category weights, preference-click cosine, and preference-click JS divergence are excluded from predictive feature sets. As label-proximity sensitivities, we refit the full model after excluding the 19 observed empty stated-category states (AUC 0.853) and after removing both profile-support count features (AUC 0.862). Dense-history checks retain the divergence threshold estimated from the full eligible cohort, which lowers positive-class prevalence after cohort restriction.

The ablation and sensitivity results are descriptive: behavioral, surface-use, profile, and exposure summaries provide diagnostic signal, but they should not be interpreted as causal mechanisms or as a production targeting model.

\section{Reproducibility Artifacts}

Public reproducibility uses metric code, aggregate tables, figure scripts, release-safe notebooks, manuscript artifacts, and synthetic schema documentation, allowing readers to inspect definitions, figures, tables, and manuscript claims without private platform traces. Authorized private reruns use output-stripped notebooks and processed logs under the same metric definitions, joins, thresholds, and model splits; raw logs, row-level click histories, profile snapshots, and production identifiers are not released.

\end{multicols}
\end{document}